\documentclass[prb,showpacs,preprintnumbers,amsmath,amssymb,twocolumn,superscriptaddress]{revtex4-1}

\usepackage[dvipdfmx]{graphicx}
\usepackage{dcolumn}
\usepackage{bm}
\usepackage{color}

\usepackage{amsmath}	
\usepackage{amsfonts}
\usepackage{braket}

\usepackage{cases}

\begin{document}

\title{
Quantum phase transition and criticality \\
in quasi one-dimensional spinless Dirac fermions
}

\author{Yasuhiro Tada}
\email[]{tada@issp.u-tokyo.ac.jp}
\affiliation{Institute for Solid State Physics, University of Tokyo, Kashiwa 277-8581, Japan}

\begin{abstract}
We study quantum criticality of spinless fermions on the quasi one dimensional 
$\pi$-flux square lattice in cylinder geometry,
by using the infinite density matrix renormalization group and abelian bosonization. 
For a series of the cylinder circumferences $L_y=4n+2=2, 6, \cdots$ with the periodic boundary condition, 
there are quantum phase transitions from gapped Dirac fermion states to charge density wave (CDW) states. 
We find that the quantum phase transitions for such circumferences are continuous and belong to the (1+1)-dimensional Ising universality class.
On the other hand, when $L_y=4n=4, 8, \cdots$, there are gapless Dirac fermions at the non-interacting point
and the phase transition to the CDW state is Gaussian.
Both of these two criticalities are described in a unified way by the bosonization.
We clarify their intimate relationship and demonstrate that a central charge $c=1/2$ Ising transition line arises as a
critical state of an emergent Majorana fermion from the $c=2$ Gaussian transition point.
\end{abstract}


\maketitle

\section{introduction}

Criticality associated with a phase transition is one of the central issues in condensed matter physics.
Various phase transitions have been established mainly for insulators which are well described by bosonic models such as Ising, XY, and Heisenberg models.
However, phase transitions in \textit{metals} where gapless fermions are coupled with bosons 
are rather poorly understood compared to insulators only with bosons.
In such a system, fermions strongly affect low energy behaviors of the bosonic order parameters and consequently could change 
criticality of the phase transition.
The critical bosonic fluctuations in turn influence the fermions, and resulting non-Fermi liquid like behaviors are often observed in various systems~\cite{MoriyaUeda2000,QCPreview2007,Brando2016,Berg2019}.

The criticality depends on structures of fermionic excitations such as dimensionality of the Fermi surface
and the number of fermion flavors (orbitals and spins).
One of the simplest examples is the spinless fermions on a one-dimensional (1D) 
chain at half-filling with the nearest neighbor 
repulsive interaction $V$, where the classical ground states for $V\rightarrow\infty$ are the charge density wave (CDW) states~\cite{Giamarchi,GNT}. 
When one introduces fermionic hopping $t$, 
there will be a Kosterlitz-Thouless phase transition to a Tomonaga-Luttinger liquid,
which is disctinct from the Ising transition in bosonic models such as the transverse Ising model.
Quantum criticality in higher dimensional systems are also of great interest, and in this context,
a semi-metallic system is an ideal platform to study interplay between fermions and bosons where the Fermi surface is a point.
Indeed, critical behaviors of phase transitions in Dirac systems have been extensively studied,
and the gapless Dirac excitations can lead to new criticalities such as chiral Ising, 
chiral XY, chiral Heisenberg universality classes
~\cite{Sorella1992,Assaad2013,Wang2014,Wang2016,Li_PRB2015,Li2015,Toldin2015,Otsuka2016,su4_2018,
Corboz2018,Rosenstein1993,Wetterich2001,Herbut2006,Herbut2009,Herbut2014,Ihrig2018}.
The critical exponents of these phase transitions have been evaluated accurately by several methods, e.g. analytical calculations and unbiased
quantum Monte Calro simulations.
In these (semi)metallic systems, the gapless fermions play essential roles
and the resulting quantum criticality is different from that in the corresponding purely bosonic system
with gapped fermions.

\begin{figure}[tbh]
\includegraphics[width=4.5cm]{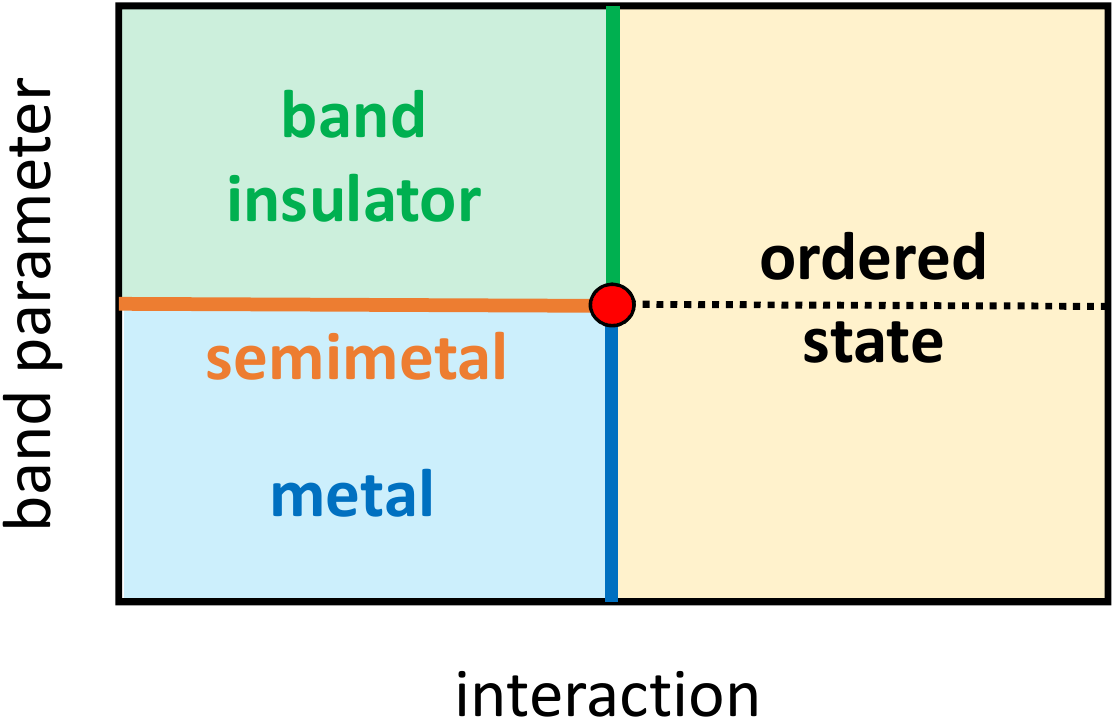}
\caption{A schematic phase diagram including both insulating and (semi)metallic states.
Generally, the blue and green phase transition lines and the red transition point 
would be characterized by different criticalities. 
}
\label{fig:phases}
\end{figure}

These two criticalities are usually studied separately
as distinct properties of metals and insulators. 
For example, the quantum phase transition from a gapless Dirac state to an antiferromagnetic state 
in a honeycomb lattice
is described by (2+1)D chiral Heisenberg universality class, while the one from a spin-orbit coupled gapped Dirac
state to the antiferromagnetic state belongs to 3D XY universality class
~\cite{Toldin2015,Lee2011,Hohenadler2012}.
Similarly, one can separately discuss two criticalities of phase transitions 
from a metal or a band insulator to an ordered state in general.
However, such separate discussions would be somewhat subtle when the band gap is very small,
and there will be crossover between fermionic criticality and bosonic criticality in a narrow gap system. 
Then, a natural question is that how these two criticalities are connected along the critical line of the phase 
transition
in an extended phase diagram including both metals and insulators (Fig.~\ref{fig:phases}).

In this study, 
we consider quasi 1D half-filled spinless fermions on a $\pi$-flux square lattice in cylinder geometry with the circumference $L_y$,
as a simple example for the quantum phase transition of $\mathbb{Z}_2$ symmetry breaking.
When the nearest neighbor repulsive interaction $V$ is weak, there are Dirac fermions with a mass $m$ due to the finite system size $L_y$
for $L_y=2,6,10,\cdots$ under the periodic boundary condition along the $y$-direction,
while there are gapless Dirac fermions at $V=0$ for $L_y=4n=4, 8, \cdots$.
The system exhibits a staggered CDW ordered state for large $V$.
The quantum phase transition is studied with use of the infinite density matrix renormalization group (iDMRG)
~\cite{White1992,DMRG_review1,DMRG_review2,DMRG_review3,TenPy1,TenPy2} together with
the recently developed scaling analysis~\cite{Corboz2018}.
Then,
we demonstrate that the quantum phase transition at a critical $V=V_c>0$ between the gapped Dirac fermions and the CDW state is continuous,
and the corresponding criticality is simply (1+1)D Ising universality class.
On the other hand, the iDMRG results suggest that the phase transition from the gapless Dirac state 
is smooth around $V=0$, which turns out to be Gaussian.
These two behaviors are well described within the bosonization approach in a unified manner, and 
a global phase diagram in the $V$-$m$ plane is discussed.
We clarify their intimate relationship and demonstrate that the central charge $c=1/2$ Ising transition line arises as a
critical state of an emergent Majorana fermion from the $c=2$ Gaussian transition point.

\section{model and phase transition}

\subsection{Model}

\begin{figure}[tbh]
\includegraphics[width=8.6cm]{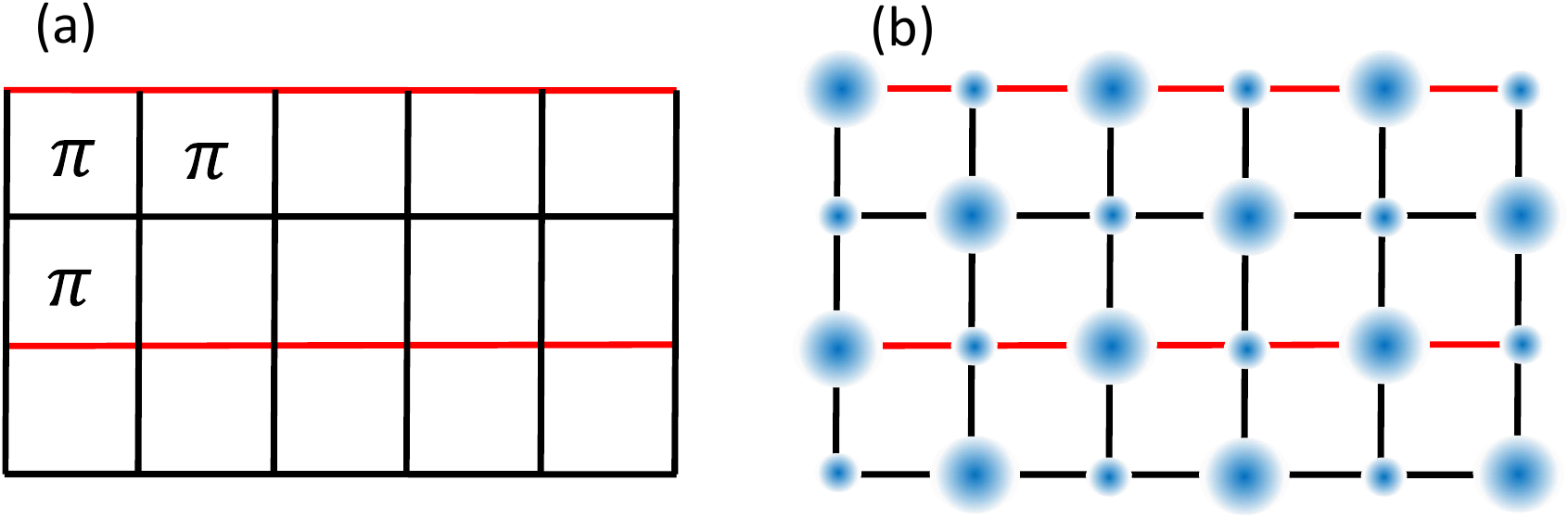}
\caption{(a) A $L_y=4$ $\pi$-flux square lattice. The hopping on the black bonds is $-t$
and that on the red bonds is $+t$, which gives a $\pi$-flux for each square plaquette.
(b) Scmematic picuture of the staggered CDW order. The blue circles represent the fermion particle density.
}
\label{fig:lattice}
\end{figure}

We consider spinless fermions on a $\pi$-flux square lattice at half-filling,
\begin{align}
H=-\sum_{\langle i,j\rangle}t_{ij}c^{\dagger}_ic_j+V\sum_{\langle i,j\rangle}n_in_j,
\end{align}
where $t_{ij}=t (-t)$ along the $x$-direction at even (odd) $y_i$ and $t_{ij}=t$ along the $y$-direction.
$\langle i,j\rangle$ represents a pair of nearest neibghbor sites (Fig.~\ref{fig:lattice}).
We use the energy unit $t=1$.
The system size is $L_x\times L_y=\infty \times L_y$ with the periodic boundary condition for the $y$-direction
otherwise specified.
In 2D ($L_y=\infty$) at $V=0$, 
this model has two Dirac points and there is a continuous 
quantum phase transition to a staggered CDW state at 
$V_c\simeq 1.30t$~\cite{Wang2014,Wang2016,Li_PRB2015,Li2015}.
The criticality of the CDW phase transition belongs to the (2+1)D chiral Ising universality class,
whose
critical exponents are evaluated as $\beta\simeq 0.60\pm 0.07$ and $\nu\simeq 0.79\sim 0.80$ by the quantum Monte Carlo calculations~\cite{Wang2014,Wang2016,Li_PRB2015,Li2015}.

\begin{figure}[tbh]
\includegraphics[width=8.6cm]{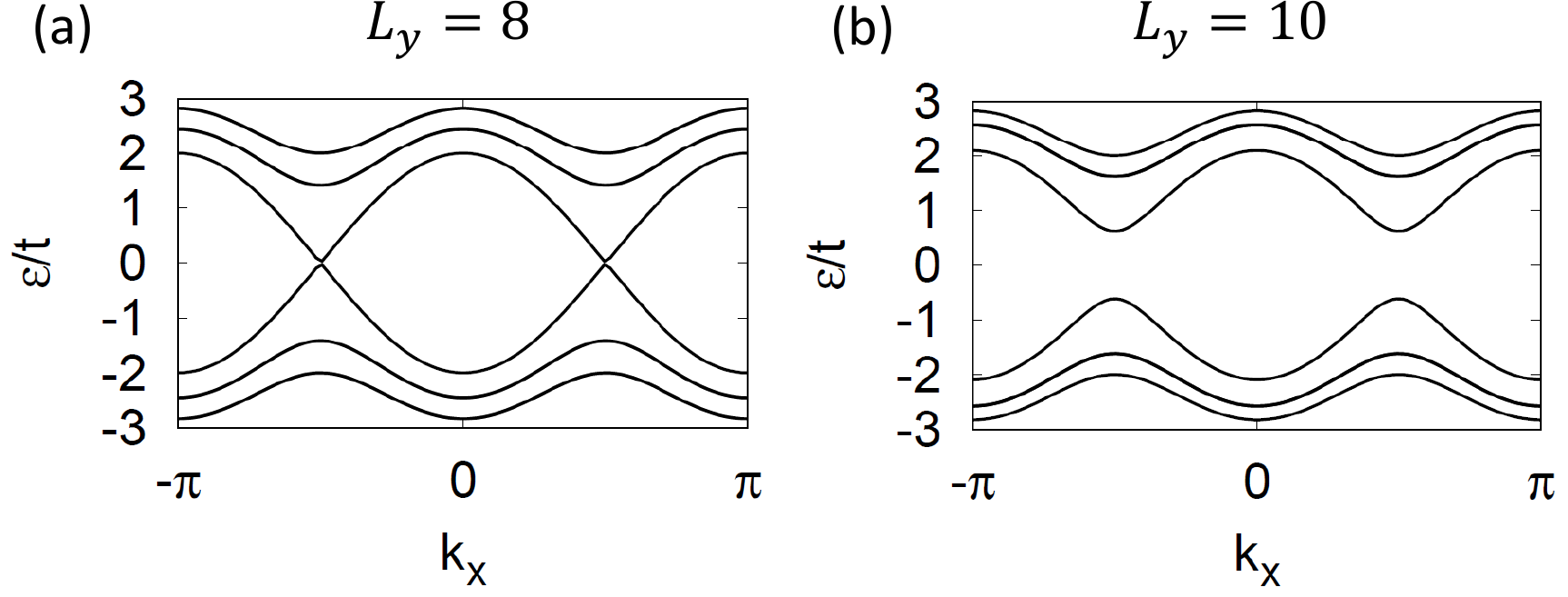}
\caption{Single particle dispersion relations (a) for $L_y=8$ 
and (b) for $L_y=10$ under the periodic boundary condition in the $y$-direction. 
}
\label{fig:dispersion}
\end{figure}

For a finite $L_y>2$,  
the single particle dispersion under the periodic boundary condition for the $y$-direction is given by 
\begin{align}
\varepsilon(k_x,k_y)=\pm \sqrt{(2t\cos k_x)^2+(2t\cos k_y)^2},
\label{eq:dispersion}
\end{align}
where $k_x$ takes continuum values and $k_y=2\pi n/L_y, (n=0,1,\cdots, L_y/2-1)$.
Similarly, $\varepsilon(k_x)=\pm \sqrt{(2t\cos k_x)^2+t^2}$ for $L_y=2$.
Due to the discreteness of $k_y$, the dispersion is qualitatively different when 
$L_y=4n=4,8,12,\cdots$ and $L_y=4n+2=2,6,10,\cdots$;
the gapless Dirac points exist for $L_y=4n$, while
the Dirac fermions are massive with the gap size $m\sim t_y/L_y$ for $L_y=4n+2$.
$\varepsilon(k)$ is shown in Fig.~\ref{fig:dispersion} for $L_y=8$ and $L_y=10$ as an example.

To discuss effects of the interaction $V$,
we use iDMRG for a system of cylinder geometry and abelian bosonization. 
The iDMRG allows a highly accurate calculation, and has been used extensively not only for one dimensional sytems
but also for two dimensional systems. 
One can directly describe a quantum phase transition of discrete symmetry in
such an infinite length cylinder by using iDMRG.
Later, we also perform bosonization analysis around $V=0$ but with a twisted boundary condition for the 
$y$-direction, which enables us to discuss the gapped and gapless fermions on an equal footing.

\subsection{iDMRG calculations}
\subsubsection{Order parameter}

\begin{figure}[tbh]
\includegraphics[width=8.6cm]{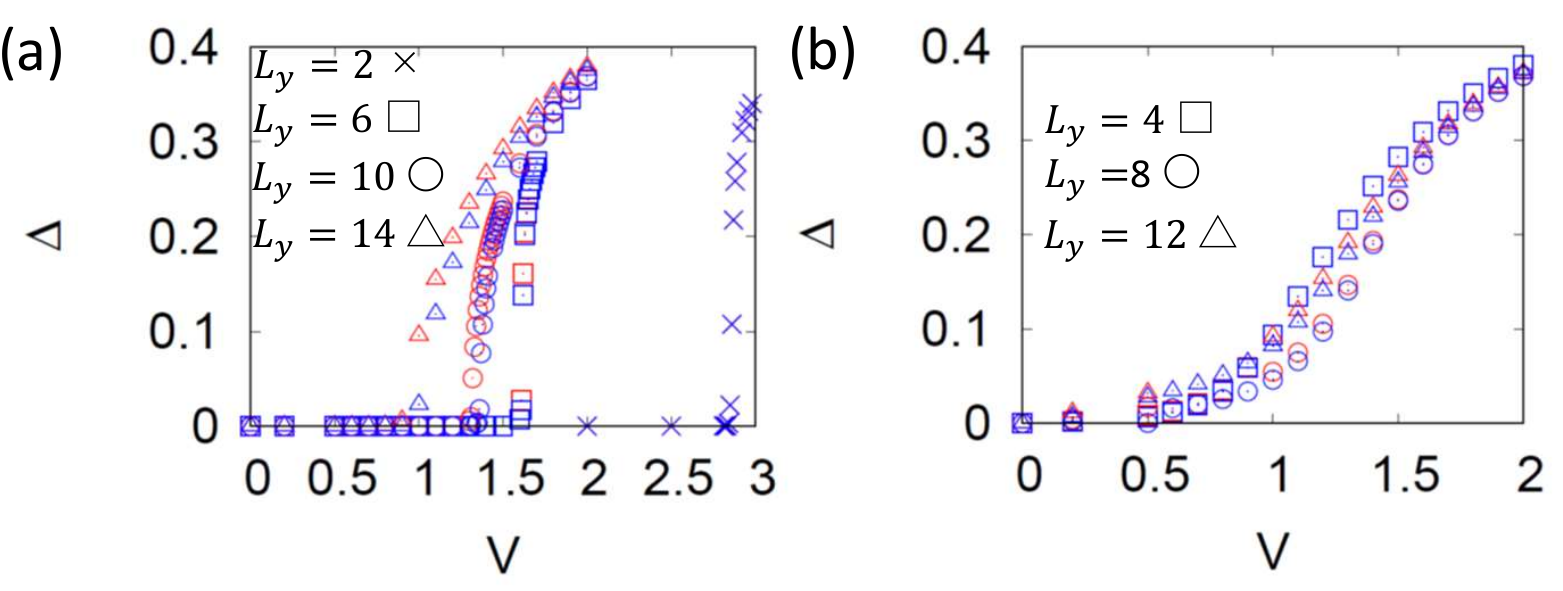}
\caption{The CDW order parameter $\Delta$ as a function of the interaction $V$ 
calculated by iDMRG with the periodic boundary condition for
the $y$-direction. 
(a) $L_y =4n+2=6, 10, 14$ with $\chi=1000$ (red), 1600 (blue). For $L_y=2$, $\chi=100$ (red), 200 (blue).
(b) $L_y =4n=4, 8, 12$ with $\chi=1000$ (red), 1600 (blue).
Note that the data for $L_y=2, 4$ with the different values of $\chi$ almost coincide in the present scale of the figures.
}
\label{fig:CDW_V}
\end{figure}

In this section, the CDW quantum phase transition is investigated by iDMRG
~\cite{White1992,DMRG_review1,DMRG_review2,DMRG_review3}
with use of the open source code TenPy~\cite{TenPy1,TenPy2}.
We discuss the CDW order parameter associated with the $\mathbb{Z}_2$ symmetry breaking,
\begin{align}
\Delta=\frac{1}{L_x'L_y}\sum_i(-1)^{|i|}n_i,
\end{align} 
where $L_x'$ is the unit period assumed in the iDMRG calculation.
The summation is over $x=1,2,\cdots,L_x'$ and $y=1,2,\cdots, L_y$.
We have performed calculations for various $L_x'$ and confirmed that the results are essentially independent of $L_x'$.
Firstly, we show $|\Delta|$ for the massive case ($L_y=4n+2$) and massless case ($L_y=4n$) respectively in 
Fig.~\ref{fig:CDW_V}.
For the massive case $L_y=2, 6, 10, 14$, we find a clear quantum phase transition from the gapped Dirac state
to the CDW state at $L_y$-dependent critical values $V=V_c(L_y)>0$.
The critical value $V_c(L_y)$ decreases as $L_y$ increases for a fixed bond dimension $\chi$,
because the Dirac band mass $m\sim t_y/L_y$ is reduced for larger $L_y$.
We expect that $V_c(L_y)$ is monotonically decreasing and approaches the 2D value $V_c(\infty)=1.30$, 
although $V_c(14)$ for $\chi$ used is smaller than $V_c(\infty)$ due to the strong finite $\chi$ effect.
On the other hand, for the massless case with $L_y=4, 8, 12$, 
the order parameter $\Delta$ behaves smoothly as a function of $V$ since the gapless Dirac states can be correctly
described only when the bond dimension $\chi$ in the iDMRG calculation is infinitely large $\chi\rightarrow\infty$.
In this limit, we expect a Gaussian transition takes place at $V=0$, 
which is indeed described by the bosonization in the later section.
In the next part, we focus on the massive case $L_y=4n+2$ and discuss its criticality within iDMRG.

\subsubsection{Finite correlation length scaling for $L_y=4n+2$}

The criticality of the phase transition for $L_y=4n+2=2 ,6, 10, \cdots$ is expected to be (1+1)D Ising universality class
if it is continuous,
because the CDW state breaks $\mathbb{Z}_2$ translation symmetry and there is no gapless Dirac fermions 
at $V=0$ for
these $L_y$.
In order to examine the criticality numerically, 
we use the scaling ansatz recently developed for tensor network states in iPEPS~\cite{Corboz2018}.
Since the one-dimensional system size $L_x$ is infinite in iDMRG, 
criticality is controlled not by $L_x=\infty$ but by the correlation length $\xi_{\chi}$
in our calculations. 
The correlation length $\xi_{\chi}$ is computed from the second largest eigenvalue of the transfer matrix for a given bond dimension $\chi$,
and $\xi_{\chi}$ characterizes finite bond dimension effects.
One would naively expect that
the system may exhibit the (2+1)D $\chi$-Ising criticality if $\xi_{\chi}\ll L_y$, while it shows (1+1)D bosonic Ising criticality if $\xi_{\chi}\gg L_y$.
In the following, we focus only on the latter case with $\xi_{\chi}\gg L_y$.

The scaling ansatz for the ground state energy density is written as
\begin{align}
E(g,h,\xi_{\chi}^{-1})=b^{-2}E(b^{y_g}g,b^{y_h}h,b\xi_{\chi}^{-1}),
\label{eq:E}
\end{align}
where $g=(V-V_c(L_y))/V_c(L_y)$ and $h$ is the conjugate field to $\Delta$.
We have assumed the dynamical critical exponent is $z=1$. 
The $L_y$-dependent critical points $V_c(L_y)$ are determined so that a scaling behavior of the order parameter
Eqs.~\eqref{eq:bn}, \eqref{eq:n} hold for larger $\xi_{\chi}$.
We obtain $V_c(L_y=2)\simeq2.8686, V_c(6)\simeq1.624$, and $V_c(10)\simeq1.50$ as will be discussed in the following.
At the critical point $g=0$, the CDW order parameter exhibits the scaling behaviors 
\begin{align}
\Delta(g=0)&\sim \xi_{\chi}^{-\beta/\nu},
\label{eq:bn}\\
\frac{\partial_g\Delta(g=0)}{\Delta(0)}&\sim \xi_{\chi}^{1/\nu},
\label{eq:n}
\end{align}
which are derived from the scaling ansatz Eq.\eqref{eq:E}.
From these two equations, we can determine the critical exponents $\beta$ and $\nu$.

\begin{figure}[htb]
\includegraphics[width=8.6cm]{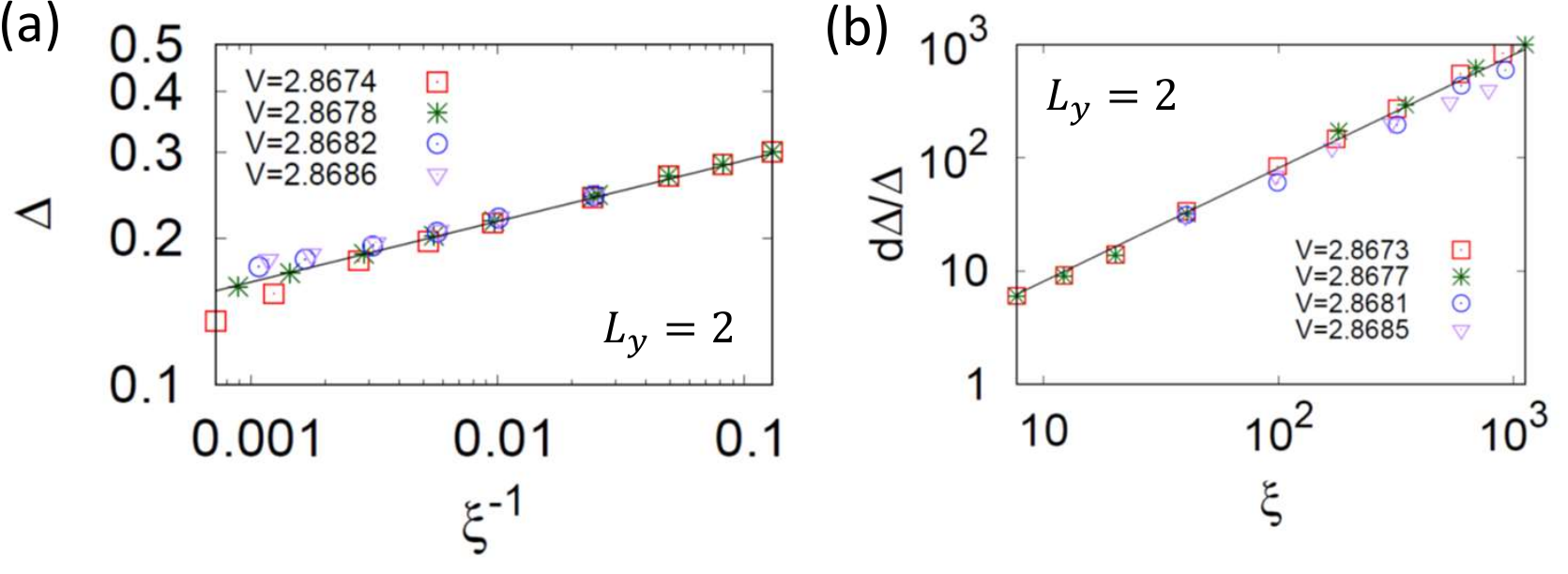}
\caption{The scaling plots of the CDW order parameter for $L_y=2$.
The correlation length $\xi_{\chi}$ is denoted as $\xi$ for simplicity.
(a) The scaling plot Eq.~\eqref{eq:bn}, and the black line is $\Delta\sim \xi^{-\beta/\nu}$ with $\beta=0.125,\nu=1$.
(b)  The scaling plot Eq.~\eqref{eq:n}, and the black line is $\partial_g\Delta/\Delta\sim \xi^{1/\nu}$ with $\nu=1$.
The $g$-derivative is approximated by $\partial_g\Delta(V)=V_c[\Delta(V+\delta V)-\Delta(V-\delta V)]/\delta V$ 
with $\delta V=0.0001$. The bond dimension is used up to $\chi\leq200$.
}
\label{fig:CDW_xi_L2}
\end{figure}

\begin{figure}[htb]
\includegraphics[width=8.6cm]{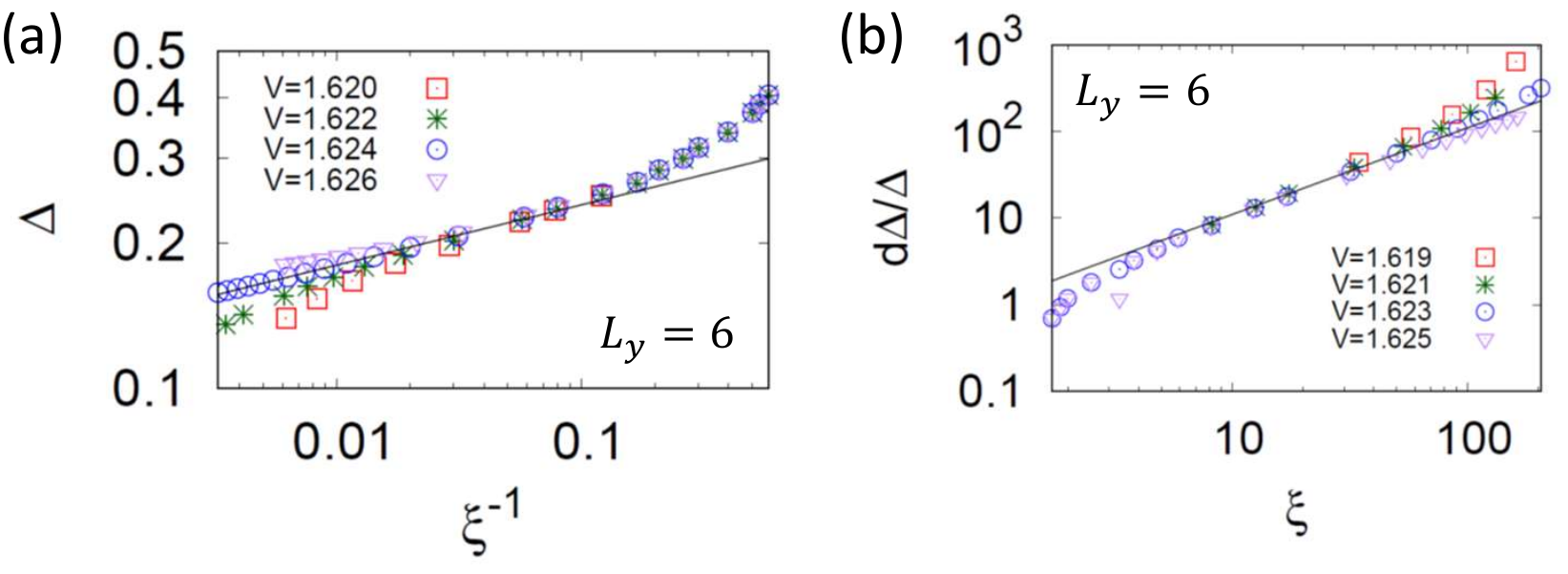}
\caption{The scaling plots of the CDW order parameter for $L_y=6$.
(a) The scaling plot Eq.~\eqref{eq:bn} and (b) Eq.~\eqref{eq:n}.
The black lines are the same as in Fig.~\ref{fig:CDW_xi_L2}, while the $g$-derivative is approximated
with $\delta V=0.001$. The bond dimension is used up to $\chi\leq2800$.
}
\label{fig:CDW_xi_L6}
\end{figure}

\begin{figure}[htb]
\includegraphics[width=8.6cm]{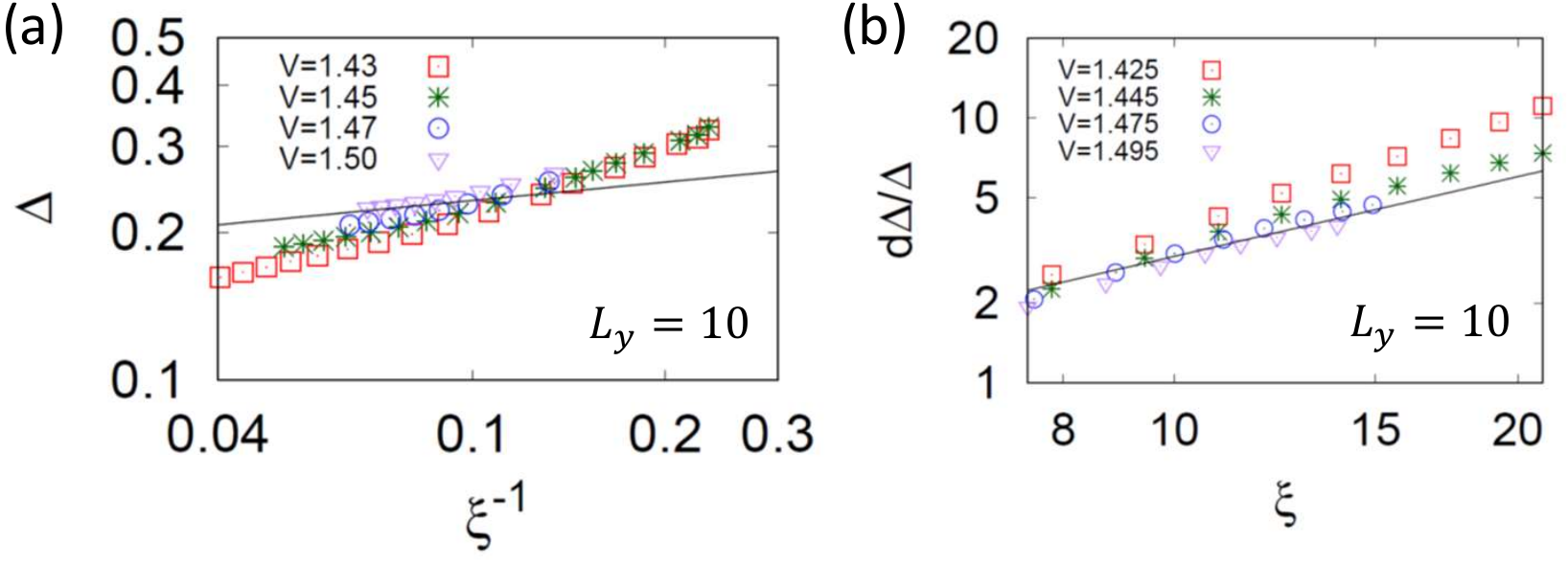}
\caption{The scaling plots of the CDW order parameter for $L_y=10$.
(a) The scaling plot Eq.~\eqref{eq:bn} and (b) Eq.~\eqref{eq:n}.
The black lines are the same as in Fig.~\ref{fig:CDW_xi_L2}, while the $g$-derivative is approximated
with $\delta V=0.005$. The bond dimension is used up to $\chi\leq2400$.
}
\label{fig:CDW_xi_L10}
\end{figure}

In Fig.~\ref{fig:CDW_xi_L2}, we show $\xi_{\chi}$-dependence of 
$\Delta$ and $\partial_g \Delta/\Delta$ for $L_y=2$.
First of all, the quantum phase transition is continuous since the scaling behaviors hold up to large $\xi_{\chi}>1000$,
although a discontinuous transition was potentially possible.
The critical interaction strength is obtained as $V_c(L_y=2)=2.8686$ from the figure.
The critical behaviors of $\Delta$ are in good agreement with those of (1+1)D Ising universality class 
with $\beta=0.125, \nu=1$, as we have expected.
Similarly, we show $\xi_{\chi}$-dependence of 
$\Delta$ and $\partial_g \Delta/\Delta$ for $L_y=6$ in Fig.~\ref{fig:CDW_xi_L6}.
The critical interaction is evaluated as $V_c(L_y=6)=1.624$.
Although there is some signature for dimensional crossover from (2+1)D chiral Ising universality class
for small $\xi_{\chi}\lesssim L_y$, the true criticality close to the critical point $g=0$ 
belongs to the (1+1)D Ising universality class.
For $L_y=10$, however, it is difficult to explicitly demonstrate the critical behavior of the (1+1)D Ising universality
class as shown in Fig.~\ref{fig:CDW_xi_L10}, because of the heavy finite $\chi$ effects.
Here, we used $\chi$ up to 2400, and the critical interaction is estimated to be $V_c(L_y=10)\simeq 1.50$.
We think that the critical behavior of the (1+1)D Ising universality class will
be reproduced for sufficiently large $\chi$ similarly to the cases for $L_y=2, 6$.

\begin{figure}[htb]
\includegraphics[width=8.6cm]{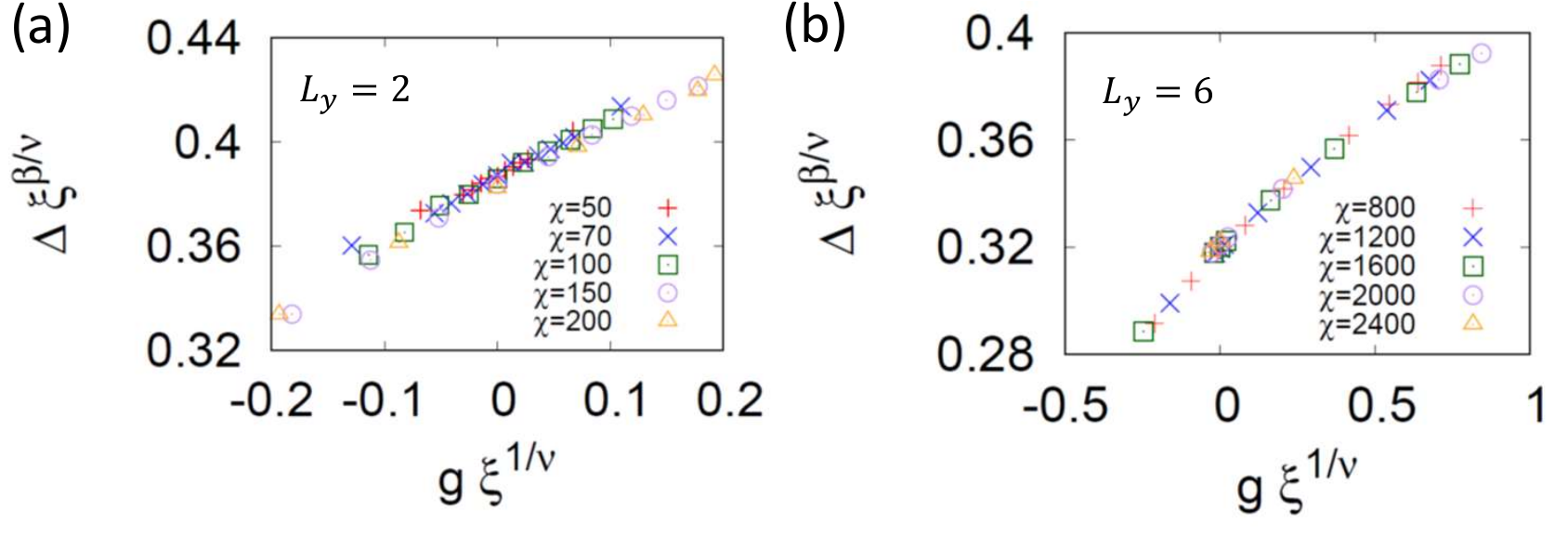}
\caption{The scaling plot of the CDW order parameter $\Delta$ for (a) $L_y=2$ and (b) $L_y=6.$
The critical exponents used are those for the (1+1)D Ising universality class $\beta=0.125,\nu=1$.
}
\label{fig:CDW_scaling}
\end{figure}

\begin{figure}[htb]
\includegraphics[width=8.6cm]{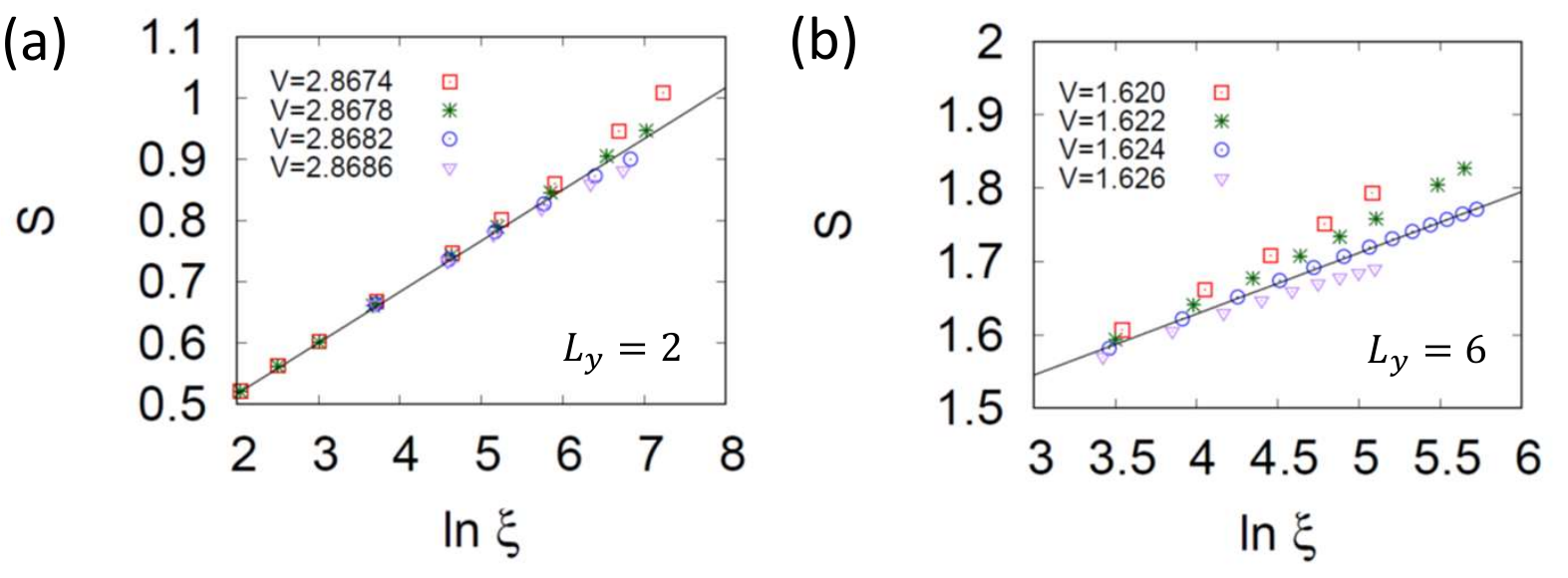}
\caption{The entanglement entropy $S$ for (a) $L_y=2$ and (b) $L_y=6.$
The black lines are $S=(c/6)\ln \xi+S_0$ with the central charge $c=1/2$.
}
\label{fig:S_EE}
\end{figure}

To further confirm the critical behaviors of the (1+1)D Ising universality class, in Fig.~\ref{fig:CDW_scaling},
we show the scaling plot
\begin{align}
\Delta \xi_{\chi}^{\beta/\nu}&= {\mathcal M}(g\xi_{\chi}^{1/\nu}),
\label{eq:mM}
\end{align}
where ${\mathcal M}$ is a scaling function.
Here, we have used only the data for $\xi_{\chi}>L_y$ to avoid effects of the dimensional crossover. 
All the data collapse into a single curve in each system size $L_y=2, 6$, which gives a cross check for 
the Ising universality class of the CDW phase transition.
Finally, Fig~\ref{fig:S_EE} shows the entanglement entropy $S$ 
for bipartitioning the infinite one dimensional chain in the iDMRG calculation
into two half-infinite chains.
In such bipartitioning, the entanglement entropy at the critical point is characterized by 
the central charge $c$ of the underlying conformal field theory and is given by
\begin{align}
S=\frac{c}{6}\ln \xi_{\chi}+S_0,
\end{align}
where $S_0$ is a constant~\cite{TenPy1,Calabrese2004}.
In the present system, the calculated $S$ at the critical point is well fitted by this formula with $c=1/2$,
which means that the corresponding conformal field theory is the $c=1/2$ Ising theory
in agreement with the critical behaviros of the order parameter $\Delta$.


\subsection{Bosonization and global phase diagram}
In this section, we discuss the relationship between the CDW phase transitions from gapless and
gapped Dirac states within the bosonization approach~\cite{Giamarchi,GNT,Balents1997,Carr2006}.
Our primary purpose is to find an effective theory description for the iDMRG calculation results.
To discuss the gapless and gapped states on an equal footing, 
we introduce the twisted boundary condition with the twist angle $\theta$ for the $y$-direction,
or equivalently insert a flux $\theta$ along the cylinder~\cite{He2017}.
When $\theta=0$ the boundary condition is realized and the non-interacting Dirac fermions are gapless for $L_y=4n$.
The band gap in Eq.~\eqref{eq:dispersion} is tuned by the twisting angle $\theta$ 
since the allowed discrete $k_y(=(2\pi n+\theta)/L_y)$ points for given 
finite $L_y$ changes as $\theta$ is varied.
For example in $L_y=4n$ case, the band gap becomes maximum at $\theta=\pi$, for which there is a
CDW phase transition from a gapped Dirac state whose criticality is (1+1)D Ising universality class.
In this way, one can smoothly connect the two extreme cases, the gapless Dirac semimetal 
and maximally gapped Dirac band insulator,
for fixed system size $L_y$.

We firstly consider the non-interacting excitation spectra in the $\pi$-flux cylinder for example with 
a fixed $L_y=4n$ under the periodic boundary condition as shown in 
Fig.~\ref{fig:dispersion} (a), and focus only on the gapless Dirac fermion branches and neglect other
gapped bands.
There are two pairs of linear dispersions with positive and negative velocities around $k_x=\pm \pi/2$. 
If we introduce a twist angle $\theta$, 
a band gap $m(\theta)$ will be induced in the pre-existing gapless Dirac bands. 
The two branches can be reproduced by an effective two leg ladder model
\begin{align}
H_\textrm{eff}&=\sum_{s=1,2}\sum_{i}-t_sc^{\dagger}_{is}c_{i+1s}-t_{\perp}\sum_ic^{\dagger}_{i1}c_{i2}+\textrm{(h.c.)}\nonumber\\
&\quad +\tilde{U}\sum_in_{i1}n_{i2}
+\tilde{V}\sum_{s=1,2}\sum_in_{is}n_{i+1s},
\end{align}
where $t_s=(-1)^{s+1}t, t_{\perp}=2t|\sin \theta|,\tilde{U}=2V/L_y, \tilde{V}=V/L_y$.
A similar effective model was studied before in the context of carbon nanotubes~\cite{Balents1997}.
By using the transformation $c_{i1}\rightarrow c_{i1},
c_{i2}\rightarrow (-1)^{i}c_{i2}$, the Hamiltonian is rewritten into the familiar form
with an additional staggered hybridization term $(-1)^it_{\perp}$,
\begin{align}
H_\textrm{eff}&\rightarrow 
\sum_{s=1,2}\sum_{i}-tc^{\dagger}_{is}c_{i+1s}-t_{\perp}\sum_i(-1)^{i}c^{\dagger}_{i1}c_{i2}+\textrm{(h.c.)}\nonumber\\
&\quad +\tilde{U}\sum_in_{i1}n_{i2}
+\tilde{V}\sum_{s=1,2}\sum_in_{is}n_{i+1s},
\end{align}
where hopping along the chain is $t$ for both $s=1,2$.

The fermion operators are approximated around the Fermi point $k_F=\pm \pi/2a$
as $\psi_s(x)=e^{-ik_Fx}\psi_{Ls}(x)+e^{ik_Fx}\psi_{Rs}(x)$ with 
$\psi_{rs}(x)=\eta_{rs}e^{-i(r\phi_{s}-\theta_s)}/\sqrt{2\pi a}$,
where $a$ is the lattice constant and $\eta_{rs}$ is the Klein factor~\cite{Giamarchi,GNT}.
The bosonic phase operators satisfy the commutation relation
\begin{align}
[\phi_{s}(x),\partial_{x'}\theta_{s'}(x')]=i\pi\delta_{ss'} \delta(x-x').
\end{align}
Furthermore, we introduce new fields $\phi_{0,\pi}=(\phi_1\pm\phi_2)/\sqrt{2}$ for convenience.
Then the Hamiltonian is bosonized into
\begin{align}
H_\textrm{eff}&=H_\textrm{kin}+H_\textrm{int},\\
H_\textrm{kin}&=\sum_{k=0,\pi}\frac{v_k}{2\pi}\int dx [K_{k}^{-1}(\partial\phi_k)^2+K_{k}(\partial \theta_k)^2], \nonumber\\
H_\textrm{int}&=\int dx [g_1\cos\sqrt{8}\phi_0+g_2\cos\sqrt{8}\phi_{\pi}\nonumber\\
&\quad+g_3\cos\sqrt{8}\phi_0\cos\sqrt{8}\phi_{\pi}
+g_4\cos \sqrt{2}\phi_{0}\sin\sqrt{2}\theta_{\pi}],\nonumber
\end{align}
where $g_1=-\tilde{U}/2\pi^2 a, g_2=\tilde{U}/2\pi^2 a, g_3=\tilde{V}/\pi^2 a, g_4=2t_{\perp}/\pi a$.
For small $\tilde{U},\tilde{V}$, the parameters are given by $v_0=v_F/K_0, v_{\pi}=v_F/K_{\pi}$, and
\begin{subequations}
\begin{align}
K_0^{-1}&=\sqrt{1+\frac{a}{\pi v_F}(\tilde{U}+4\tilde{V})}\simeq 1+\frac{a}{2\pi v_F}(\tilde{U}+4\tilde{V}), \\
K_{\pi}^{-1}&=\sqrt{1+\frac{a}{\pi v_F}(-\tilde{U}+4\tilde{V})}\simeq 1+\frac{a}{2\pi v_F}(-\tilde{U}+4\tilde{V}),
\end{align}
\end{subequations}
where $v_F=2t$ is the Fermi velocity of the non-interacting model.
The scaling dimensions of the operators are easily read off as
\begin{subequations}
\begin{align}
[g_1]&=2K_0\simeq 2-\frac{a}{\pi v_F}(\tilde{U}+4\tilde{V}), \\
[g_2]&=2K_{\pi}\simeq 2-\frac{a}{\pi v_F}(-\tilde{U}+4\tilde{V}), \\
[g_3]&=2K_0+2K_{\pi}\simeq 4-\frac{8a}{\pi v_F}\tilde{V}, \\
[g_4]&=\frac{K_0}{2}+\frac{1}{2K_{\pi}}\simeq 1-\frac{a}{2\pi v_F}\tilde{U}. 
\end{align}
\end{subequations}

We first consider the case with $t_{\perp}=0$, or equivalently $g_4=0$.
Then, the most relevant term is the $g_1$-term, and 
$\phi_0$-field gets pinned to $\langle\phi_0\rangle=0$ because of the strong coupling $g_1\rightarrow-\infty$.
The remaining $g_2,g_3$-terms will have the same functional form, $\cos\sqrt{8}\phi_{\pi}$,
and be renormalized to $g_2,g_3\rightarrow\infty$.
Therefore, both of the two fields $\phi_0,\phi_{\pi}$ become gapped as long as $V>0$,
and the phase transition is a Gaussian transition from the $c=2$ two-flavor gapless Dirac state
to the fully gapped CDW state.
This is consistent with the iDMRG calculation where the CDW order parameter is non-zero 
for very small $V$ when $L_y=4n=4,8,\cdots$ under the periodic boundary condition.

Next, we consider a very small $0<t_{\perp}\ll V$, for which the renormalized parameters still satisfy 
$|g_4|\ll |g_1|$ down to some energy scale under the renormalization group.
In this energy scale, $\phi_0$-field is nearly locked as $\langle\phi_0\rangle\simeq 0$ and the low energy physics is described by the $\phi_{\pi}$-field only,
\begin{align}
H_\textrm{eff}&\simeq 
\frac{v_{\pi}}{2\pi}\int dx [K_{\pi}^{-1}(\partial\phi_{\pi})^2+K_{\pi}(\partial \theta_{\pi})^2]\nonumber\\
&\quad +\int dx [g_{23}\cos \sqrt{8}\phi_{\pi}+g_4\sin \sqrt{2}\theta_{\pi}],
\label{eq:H_dsg_}
\end{align}
where $g_{23}=g_2+g_3$ and we have used the approximation $\langle\cos\sqrt{8}\phi_0\rangle
\simeq \langle\cos\sqrt{2}\phi_0\rangle\simeq1$.
Note that the parameters in Eq.~\eqref{eq:H_dsg_} should be regarded as renormalized ones under the 
renormalization group flow down to the above mentioned energy scale.
In this Hamiltonian, the $g_{23}$-term favors the CDW state while the $g_4$-term leads to the band insulator,
and this competition can lead to a gapless state when these two perturbations cancel each other.
The resulting gapless state is described by the $c=1/2$ Majorana fermions, which corresponds to the
criticality of the CDW phase transition from the band gapped Dirac state discussed in the previous section.
To see this, we focus on a fine-tuned state where the two perturbation terms 
are maximally competing having the same scaling dimensions,
$[g_{23}]=[g_4]$, namely
\begin{align}
2K_{\pi}=\frac{1}{2K_{\pi}} \Rightarrow K_{\pi}=\frac{1}{2}.
\end{align}
By redefining the boson fields as $\phi'_{\pi}=\phi_{\pi}/\sqrt{K_{\pi}}, \theta'_{\pi}=\sqrt{K_{\pi}}\theta_{\pi}-\pi/4$
with $K_{\pi}=1/2$,
the Hamiltonian is rewritten as
\begin{align}
H_\textrm{eff}&=\frac{v_{\pi}}{2\pi}\int dx [(\partial\phi_{\pi}')^2+(\partial \theta_{\pi}')^2]\nonumber\\
&\quad +\int dx [g_{23}\cos 2\phi_{\pi}'+g_4\cos 2\theta_{\pi}'].
\end{align}
This Hamiltonian is called the self-dual sine-Gordon model and has been studied extensively
~\cite{GNT,Shelton1996,Shelton1996_2,dsG2002,Robinson2019}.
Since the scaling dimensions of both $g_{23},g_4$-terms are 1,
one can refermionize them by using  a spinless fermion operator $\psi_r(x)\simeq 
\eta_{r}e^{-i(r\phi'_{\pi}-\theta_{\pi}')}
/\sqrt{2\pi a}$ as
\begin{subequations}
\begin{align}
\cos2\phi_{\pi}'&= -i\pi a[\psi_R^{\dagger}\psi_L-\psi_L^{\dagger}\psi_R],\\
\cos2\theta_{\pi}'&=-i\pi a[\psi_R^{\dagger}\psi_L^{\dagger}-\psi_L\psi_R].
\end{align}
\end{subequations}
Therefore the self-dual sine-Gordon model is mapped to a free spinless fermion model with mass terms,
\begin{align}
H_\textrm{eff}&=\int dx -iv_{\pi}[\psi_R^{\dagger}\partial\psi_R-\psi_L^{\dagger}\partial\psi_L]\nonumber\\
&\quad -im_{23}[\psi_R^{\dagger}\psi_L-\psi_L^{\dagger}\psi_R]
-im_4[\psi_R^{\dagger}\psi_L^{\dagger}-\psi_L\psi_R],
\end{align}
where $m_{23}=\pi a g_{23}, m_4=\pi a g_4$
Then we introduce Majorana fermions $\gamma^1=(\psi+\psi^{\dagger})/\sqrt{2},
\gamma^2=(\psi-\psi^{\dagger})/\sqrt{2}i$ to write the Hamiltonian in the Majorana basis,
\begin{align}
H_\textrm{eff}&=\sum_{a=1,2}\int dx -i\frac{v_{\pi}}{2}[\gamma_R^{a}\partial\gamma_R^a
-\gamma_L^{a}\partial\gamma_L^a]-im_{\gamma a}\gamma_R^a\gamma_L^a,
\end{align}
where $m_{\gamma1}=m_{23}+m_4, m_{\gamma2}=m_{23}-m_4$.
Clearly, only one Majorana fermion $\gamma_2$ is gappless and the other one $\gamma_1$ is gapped
along the special line given by $m_{23}=m_4$ in the $V$-$t_{\perp}$ plane.
(Note that we have assumed $t_{\perp}>0$ and thus $m_{\gamma1}\neq0$ in this study.)
This emergent gapless Majorana fermions describe the $c=1/2$ conformal field theory 
which is the critical theory for the CDW phase transition from the band gapped Dirac state
studied in the previous section.
Physically, the Majorana fermions correspond to domain walls of the CDW oder.

\begin{figure}[htb]
\includegraphics[width=8.6cm]{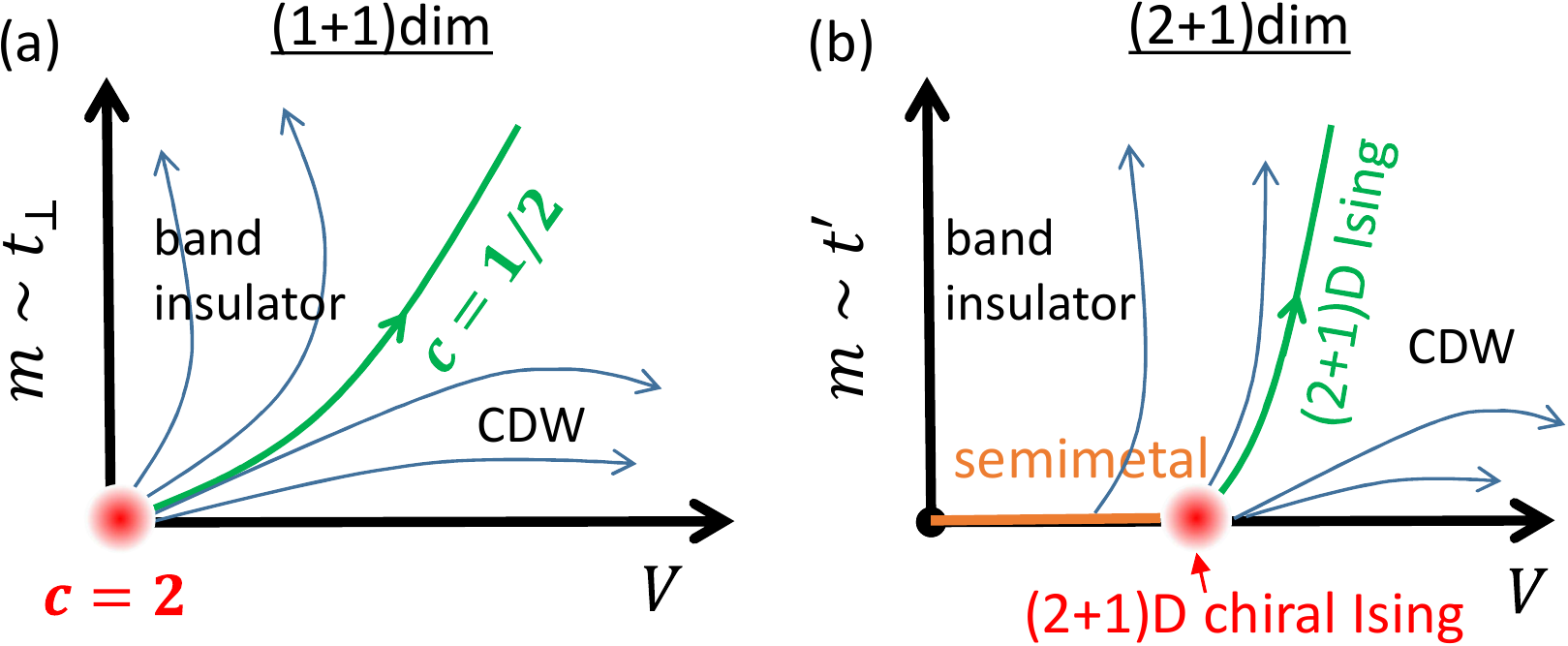}
\caption{(a) Schematic global phase diagram in the $V$-$t_{\perp}$ plane in (1+1)D.
The red point at the origin is the $c=2$ Gaussian transition point,
and the green curve is the $c=1/2$ Ising transition line separating the band insulator and CDW state.
The arrows correspond to the renormalization group flow in the effective low energy model.
(b) Expected phase diagram in (2+1)D.
$t'$ is an additional hopping which induces a band gap.
}
\label{fig:phase_diagram}
\end{figure}

We have shown within the bosonization how the fermionic criticality at the Gaussian transition
is connected to the bosonic criticality at the Ising transition.
These discussions are summarized in the global phase diagram shown in Fig.~\ref{fig:phase_diagram}.
We expect that 
competition between the band gap and interaction would be important also for higher dimensions.
For example in spinless fermions on the two dimensional $\pi$-flux square lattice,
there is a CDW quantum phase transition with (2+1)D chiral Ising criticality at $V=V_c>0$ from the
gapless Dirac semimetal~\cite{Wang2014,Wang2016,Li_PRB2015,Li2015}, 
while a transition from the gapped Dirac insulator is expected to show 3D Ising criticality if it is continuous.
The two phase transitions would be connected in a non-trivial way,
and the familiar 3D Ising criticality might be understood as a critical state of an emergent
object 
from the (2+1)D chiral Ising critical point.
Further studies are necessary to develop theoretical understanding of these issues.

\section{summary and discussion}

We have studied the CDW quantum phase transition and its criticality in spinless fermions
on the quasi one dimensional $\pi$-flux square lattice, by using iDMRG and bosonization.
We find that the phase transition from a Dirac band insulator is continuous and its universality class is 
(1+1)D Ising with the central charge $c=1/2$ when $L_y=4n+2=2, 6, \cdots$ under the periodic  boundary condition, 
while
that from a Dirac semimetal is Gaussian with $c=2$ when $L_y=4n=4, 8, \cdots$.
By introducing the twisted boundary condition, we discussed how the fermionic criticality of the 
Gaussian transition in the gapless Dirac semimetal 
is connected to the bosonic criticality of the Ising transition in the gapped  Dirac band insulator.
The global phase diagram was discussed, where the $c=2$ critical point is connected to the $c=1/2$
critical line.
The resulting $c=1/2$ critical line arises from the competition between the band mass and 
the density interaction leading to the CDW gap, and is described by the emergent Majorana fermions
which is regarded as a fractionalized object.
This could give a new insight for a comprehensive understanding of phase transitions in both metals and insulators.
Our results could provide a basis to understand higher dimensional systems,
and also may be directly relevant for the artificially created $\pi$-flux systems in cold atoms with the
synthetic magnetic field~\cite{synthetic2011,Ozawa2019}.

\section*{Acknowledgement}
We are grateful to Y. Fuji for valuable discussions and constructive comments on our manuscript.
We also thank 
J. -H. Chen, R. Kaneko, M. Nakamura, M. Oshikawa, S. Takayoshi and Y. Yao for fruitful discussions.
The numerical calculations have been done at Max Planck Institute for the Physics of Complex Systems.
This work was supproted by Grants-in-Aid for Scientific Research
No. JP17K14333 and KAKENHI on Innovative Areas ``J-Physics'' [No. JP18H04318].

%

\end{document}